\documentclass[12pt]{iopart}

\usepackage{graphicx}

\begin{document}

\title{Strangeness in Astrophysics and Cosmology}

\author{T Boeckel\dag, 
M Hempel\dag,
I Sagert\ddag,
G Pagliara\dag, 
B Sa'd\dag,
J~Schaffner-Bielich\dag\footnote[3]{invited talk given at the
    International Conference on Strangeness in Quark Matter (SQM2009),
    Buzios, Brasil, September 28 -- October 2, 2009.}}

\address{
\dag\ Institut f\"ur Theoretische Physik,
Ruprecht-Karls-Universit\"at, Philosophenweg 16, 
69120 Heidelberg, Germany\\
\ddag\ Institut f\"ur Theoretische Physik, Goethe Universit\"at, 
Max-von-Laue-Str.~1, 60438~Frankfurt am Main, Germany}

\ead{schaffner@thphys.uni-heidelberg.de}

\begin{abstract}
  Some recent developments concerning the role of strange quark matter
  for astrophysical systems and the QCD phase transition in the early
  universe are addressed. Causality constraints of the soft nuclear
  equation of state as extracted from subthreshold kaon production in
  heavy-ion collisions are used to derive an upper mass limit for
  compact stars. The interplay between the viscosity of strange quark
  matter and the gravitational wave emission from rotation-powered
  pulsars are outlined. The flux of strange quark matter nuggets in
  cosmic rays is put in perspective with a detailed numerical
  investigation of the merger of two strange stars. Finally, we
  discuss a novel scenario for the QCD phase transition in the early
  universe, which allows for a small inflationary period due to a
  pronounced first order phase transition at large baryochemical
  potential.
\end{abstract}



\section{Introduction}


There is no attempt in this contribution to discuss all the recent
advances in the very active research area of studying the impacts of
strange quark matter in astrophysics and cosmology. We refer to the
other contributions of this conference for topics not covered in this
contribution to the proceedings.
Signals for the QCD phase transition to strange quark matter in
core-collapse supernovae are addressed in the contribution by Irina
Sagert et al. (see also \cite{Sagert:2008ka}). For the appearance of a
new strange phase in the evolution of proto-neutron stars we refer to
the contribution by Giuseppe Pagliara et al. (see also
\cite{Pagliara:2009dg}). The issues of how to nucleate strange quark
matter in astrophysical systems has been studied in detail in the
contribution of Bruno Mintz et al. (see also \cite{Mintz:2009ay}).
Strange quark matter, colour superconductivity and the relation to
quarkyonic matter are discussed in the contribution by David Blaschke
et al. (see also \cite{Blaschke:2008br}).

In the following we give attention to a few other recent developments
in the field. 
First we reconsider the maximum mass constraint for neutron stars
using causality arguments for the nuclear equation of state.
Constraints on the nuclear equation of state using heavy-ion data,
here subthreshold production of kaons, can be adopted to give a new
maximum mass limit of about 2.7 solar masses. Then we are studying the
gravitational wave emission from rotating neutron stars and the
relation to the viscosity of strange quark matter. The recent
simulation of the merger of two strange stars, the release of strange
quark matter nuggets to the interstellar medium and the possible
impact for the flux of strangelets in cosmic rays are another issue
being highlighted. Finally, we are concerned with the cosmological
QCD phase transition which happens around $10^{-5}$ seconds after the
big bang. If one allows for an inflationary period at a strong first
order phase transition, the early universe can actually pass through
the QCD phase diagram at large baryochemical potential in contrast to
the standard model of cosmology.


\section{Maximum mass of neutron stars:
constraints from heavy-ion data}


Neutron stars are produced in core-collapse supernova explosions and
are compact, massive objects with radii of 10--15 km and masses of
1--2 $M_\odot$. Pulsars are rotation-powered neutron stars and have
been observed in binary systems, i.e.\ with a companion star, with a
white dwarf or even with another neutron star. To create such a binary
pulsar one starts with two ordinary stars, one with at least
$8M_\odot$. The heavier star explodes first in a supernova and leaves
a neutron star as a remnant. The neutron star is spun up by accreting
matter from the companion star. Then also the companion star might
become a white dwarf or another neutron star depending on its initial mass
(see \cite{Lorimer:2008se} for a review).

For binary pulsars masses can be pretty nicely determined.  At present
more than 1800 pulsars are known with 140 binary pulsars.  From those
binary pulsars, the pulsar PSR J1903+0327 is particular interesting
\cite{Champion:2008,Freire:2009fn}. By measuring the post-Keplerian parameters from
pulsar timing, here the Shapiro delay parameters $r$ and $s$ alone,
constrain the mass to $M=(1.67\pm 0.11)M_\odot$. Combined with the
periastron advance $\dot \omega$ the mass is constraint to be
$M=(1.67\pm 0.01)M_\odot$. This mass measurement would put some severe
constraints on soft nuclear equations of state which give a smaller
maximum mass \cite{Lattimer:2006xb}.

Kaons are produced by associated production: NN$\to{\rm N}\Lambda$K
and NN$\to$NNK$\overline{\rm K}$ in elementary collisions. In the
medium as generated in central heavy-ion collisions new processes
emerge through rescattering as $\pi{\rm N}\to\Lambda$K,
$\pi\Lambda\to{\rm N}\overline{\rm K}$ so that kaons can be even
produced below the (elementary) threshold bombarding energy.  In
heavy-ion collisions of 1--2~AGeV nuclear matter can compressed up to
$3n_0$.  As kaons have a long mean-free path, they can escape the high
density region serving as a messenger of the properties of
high-density nuclear matter (for a review see
e.g.\cite{Aichelin:2008}). One defines the double ratio as the
multiplicity of produced kaons per mass number for Au+Au collisions
relative to the one for C+C collisions. This observable turns out to
be rather insensitive to input parameters but strongly dependent on
the nuclear equation of state (to be more precise the in-medium
potential of nucleons which can be related to the nuclear equation of
state). Using a simple Skyrme parametrization, only simulations with a
compression modulus of $K_N\approx 200$ MeV
\cite{Fuchs:2000kp,Hartnack:2005tr} can describe the energy dependence
of the kaon production data as measured by the KaoS collaboration
\cite{Sturm:2000dm,Forster:2007qk}.  Hence, the nuclear equation of
state would be soft around 2--3~$n_0$.

Causality arguments can then be adopted to constrain the maximum
possible mass of neutron stars.  Above a fiducial density as
determined from the data analysis of the KaoS data the stiffest
possible equation of state is taken which gives the maximum pressure
for a given energy density allowed.  Causality demands then that at
most $p=\epsilon-\epsilon_c$ above the fiducial density $\epsilon_f$
(see \cite{Rhoades:1974fn,Kalogera:1996}).  There is a general scaling
relation for the maximum possible mass of compact stars which is given by
$M_{\rm max}= 4.2M_\odot (\epsilon_0/\epsilon_f)^{1/2}$ where
$\epsilon_0$ is the energy density of normal nuclear matter (the
numerical prefactor depends slightly on the low-density nuclear
equation of state, see \cite{Hartle78}). For a fiducial density of
$\epsilon_f=2\epsilon_0$ as provided by the heavy-ion data the upper
mass limit is about $2.7 M_\odot$ \cite{Sagert:2007nt,Sagert:2007kx}.


\section{Gravitational wave emission from rotating neutron stars: 
the viscosity of strange quark matter}


There exists a special class of binary neutron stars, so called x-ray
burster. In these astrophysical systems, the neutron star is accreting
matter from an ordinary star or a white dwarf. A low-mass x-ray binary
(LMXB) is a neutron star with a low mass companion star or a white dwarf.
The accreting matter falls onto the neutron star surface and thereby
releases x-rays in a burst. The neutron star surface is heated up to
typical temperatures of $T=10-100$~keV.

For rotating neutron stars, there exists an instability, the r-mode
instability. Oscillations brings the neutron star matter out of
$\beta$-equilibrium which is restored by weak processes affecting the
bulk and shear viscosity \cite{Andersson:1997xt}. The viscosity of
nuclear matter is quite low in certain regions of frequency and
temperature so that the oscillations will not be damped, gravitational
waves are emitted and the star's rotation slows down. The
contributions from quarks increase the stability window in temperature
and frequencies substantially.  In color-superconducting phases the
viscosity of quark matter can be drastically different
\cite{Madsen:1999ci,Drago:2003wg,Jaikumar:2008kh,Mannarelli:2008je,Alford:2008pb}.
In the 2SC phase, with the pairing of two quark flavours and two
colours, there are still unpaired quarks left and one obtains similar
results compared to the free case \cite{Sa'd:2007ud,Sa'd:2008gf}. Both
cases are compatible with the data on low-mass x-ray binaries.  On the
contrary, quarks of all color and flavor are paired in the CFL
phase. The reaction rates are suppressed, the instability window is
considerably broader which could leave an observable impact on the
evolution of young pulsars and accreting neutron stars. Recently, it
was also found that there is a new process which takes away the energy
from the r-mode oscillations by the emission of neutrinos. It acts
effectively as a viscosity which would help in damping the r-mode
instability by a substantial factor \cite{Sa'd:2009vx}.


\section{Strange star mergers: the flux of strangelets in cosmic rays}


According to the Bodmer-Witten hypothesis
absolutely stable strange quark
matter would be more bound than ordinary nuclear matter.  If one
allows for absolutely stable strange quark matter, then strange stars
will exist which are compact stars consisting of absolutely stable
strange quark matter being bound by strong interactions (selfbound)
and not by gravity as ordinary neutron stars. Strange stars can
coexist with neutron stars, but in principle a neutron star would
collapse to strange star in a spectacular astrophysical event.
Recently, the collisions of two strange stars have been simulated in 
3D relativistic smoothed particle hydrodynamics with approximate
treatment of effects from General Relativity
\cite{Bauswein:2008gx}. Typical timescales involved are milliseconds,
so that matter is in $\beta$-equilibrium.  The simulations have been
performed for two bag constants of $B=60$ and 80 MeV/fm$^{-3}$, in both
cases one gets absolutely stable strange quark matter, and for
different initial masses of the two strange stars.  In principle, two
different scenarios have been found: either there is an immediate
collapse to a black hole or a hypermassive object stabilized by
differential rotation is formed. The implications of these results are
quite striking: the ejected mass of strange quark matter (strangelets)
released to the interstellar medium are estimated to be $(1.4-
2.8)\cdot 10^{-4}M_\odot$ for the lower bag constant and zero for the
larger one. As strange star mergers are considered to be the prime
source for strangelets in cosmic rays, the latter result opens the
possibility that no strangelets are seen in cosmic rays although
strange stars exist. The Alpha Magnetic Spectrometer AMS will measure
the strangelet flux in cosmic rays (planned to be installed at the
International Space Station ISS in 2010). Interestingly, the
gravitational wave pattern of compact star mergers can be used to
distinguish between neutron star and strange star mergers
\cite{Bauswein:2009im}.


\section{Early universe: phase transition from strange quark matter}


Let us briefly outline the history of the early universe.  In a
radiation dominated universe the temperature scales inversely with the
scale parameter $a$. Big-bang nucleosynthesis happens at times of $t=
1$s to 3 minutes which corresponds to temperatures of $T=1$ to
$0.1$~MeV. The QCD phase transition occurs at a temperature of about
$T\approx 170$~MeV, i.e.\ at a time of $t\approx 10^{-5}$~s after the
big bang. The early universe passes through the electroweak phase
transition at $t\approx 10^{-10}$~s ($T\approx 100$~GeV).
In the standard cosmology, one knows from the measurement of the
microwave background radiation with WMAP and from the observed light
element abundance from the big bang nucleosynthesis that the ratio of
the number of baryons to the number of photons (which is related to
the total entropy of the universe) is tiny:
\[ n_B/s \sim n_B/n_\gamma \sim \mu/T \sim 10^{-9} \] 
Note that the baryon number per entropy is a conserved quantity
throughout the evolution of the universe unless there is a
non-equilibrium process as a first order phase transition.  From the
above considerations one concludes that the early universe evolves
along $\mu \sim 0$ which implies that the early universes crosses the
QCD phase transition in a region of the QCD phase diagram where one
expects a crossover transition as seen in recent lattice gauge
calculations. A crossover transition implies that there is nothing
spectacular happening, as one expects no cosmological signals from
such a kind of transition.

The Friedmann equation for radiation dominated universe reads
\[ H^2 = \frac{8\pi G}{3} \rho \sim g(T) \frac{T^4}{M_p^2} \] where
$g(T)$ is effective number of relativistic degrees of freedom at a
given temperature $T$ and $M_p$ is the Planck mass.  The Hubble time
is defined by the Hubble parameter and is related to the true time by
$t=3t_H$ for a radiation dominated universe. By using the Friedmann
equation one finds that $t_H = 1/H \sim g^{-1/2} M_P/T^2$ so that
there is a simple relation between the time after big bang and the
temperature of the radiation dominated early universe
\[ \frac{t}{\mbox{1 sec}} \sim \left(\frac{\mbox{1 MeV}}{T} \right)^2 \]
We see that if the number of degrees of freedom changes smoothly with
the temperature, as it would be the case for a crossover phase
transition, the temperature changes continuously with time.

The interesting question is what would happen if the early universe
passes through a first order phase transition? We argue in the
following that this is a viable scenario for the QCD phase transition
and not in contradiction to our present knowledge of QCD. Furthermore,
we demonstrate that it would leave an observable imprint on the
gravitational wave background \cite{Boeckel:2009ej}.

\begin{figure}
\centering
\includegraphics[width=0.5\textwidth]{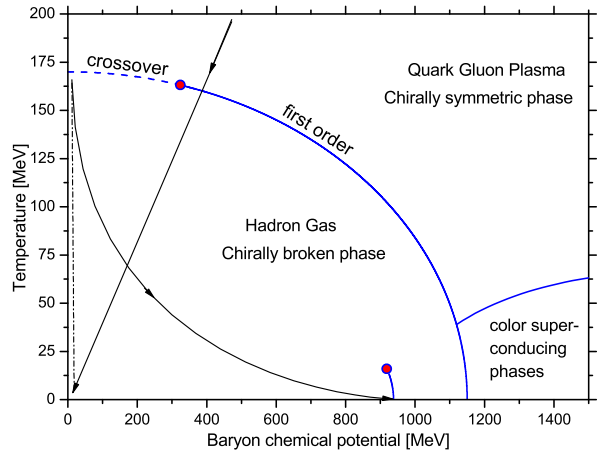}
\caption{The evolution through the QCD phase diagram for the little
  inflation scenario. The universe is trapped in a metastable vacuum
  at the first order phase transition line, supercools and is reheated
  back close to the critical temperature at tiny values of the
  baryochemical potential.}
\label{fig:evolution}
\end{figure}

A first order phase transition in general allows for a false
metastable vacuum state where the universe could be trapped for some
time.  For a constant vacuum energy, one arrives at the de Sitter
solution of the Friedmann equations which results in an (additional
small) inflationary period:
\[ H=\dot a/a \sim M_p^{-1}\rho_{\rm v}^{1/2} = H_{\rm v} = const. 
\to a \sim \exp (H_{\rm v}\cdot t) \]
Remarkably, just a few e-folds are enough (standard inflation needs
$N\sim 50$) for our purposes. The baryon to photon ratios before and
after the inflationary period should scale with the ratio of the scale
parameters cubed as entropy is produced during reheating while baryon
number is conserved so that
\[ \left(\frac{\mu}{T}\right)_f \approx \left(\frac{a_i}{a_f}\right)^3
\left( \frac{\mu}{T}\right)_i \] 
The final ratio should be $10^{-9}$ as observed so that we need just a
boost of $N = \ln \left(a_f/a_i\right) \sim \ln (10^3) \sim 7$, i.e.\
seven e-folds, to get an initial ratio of $\left(\mu/T\right)_i \sim
\cal{O}$(1). Note that for standard inflation one needs about $N=50$
e-folds, so that the little inflation at the QCD phase transition can
not replace it but could be present as an additional small
inflationary period in the early universe.

So our scenario is as follows: the early universe is initially at
large baryochemical potentials $\mu/T\sim 1$. Such a large value is
possible for e.g.\ Affleck-Dine baryogenesis which involves scalar
supersymmetric fields carrying baryon number. The early universe
reaches the first order phase transition line of QCD at high
baryochemical potentials and is trapped in the false vacuum. Now the
inflationary period starts with supercooling and dilution with $\mu/T
= const.$. The decay to the true vacuum state will release latent heat
so that the universe is reheated to $T\sim T_c$. Afterwards, the final
baryon to photon ratio is given by $\mu/T\sim 10^{-9}$ due to the
entropy produced during the transition. Finally, the universe evolves
along the standard cosmological path to big-bang nucleosynthesis and
so on. The path through the QCD phase diagram is depicted in
Fig.~\ref{fig:evolution}. 

\begin{figure}
\centering
\includegraphics[width=0.6\textwidth]{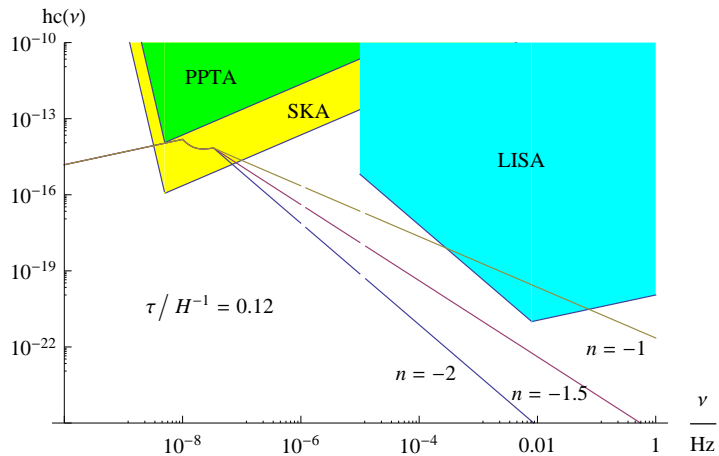}
\caption{The gravitational wave spectrum from a first order QCD phase
  transition. The Parkes Pulsar Timing Array PPTA already gives limits
  on the maximum amplitude of those gravitational waves.  In the future
  the Square Kilometre Array SKA will push these limits down by a few
  orders of magnitude. For a flat spectrum at high frequencies, LISA
  would be also sensitive.}
\label{fig:gw_spectrum}
\end{figure}

However, there are observable differences to the standard model of
cosmology. The first order phase transition produces perturbations,
scalar and tensor ones, by colliding bubbles.  The tensor modes
correspond to gravitational waves with a frequency scale given by the
(redshifted) horizon scale at the transition point
\[ \nu_{\rm peak} \sim H\cdot T_{\gamma,0}/T_{\rm QCD} \sim T_{\rm QCD}/M_p \cdot
T_{\gamma,0} \sim 10^{-7}~\mbox{\rm Hz} \] 
The maximum amplitude of the gravitational waves is then given by the
ratio of the scale factors today and at the QCD phase transition and
is $h\sim a_{\rm QCD}/a_0 \sim 10^{-12}$. For superhorizon modes $\nu
< H$ the amplitude scales as $h(\nu) \propto \nu^{-1/2}$ which
corresponds to white noise. The modes within the horizon $\nu > H$ are
determined by multi bubble collisions and could scale as $h(\nu)
\propto \nu^{n}$ where n is between $-1$ and $-2$
\cite{Kamionkowski:1993fg,Huber:2008hg}. The gravitational waves from
a first-order QCD phase transition are observable with pulsar timing,
as with the Parkes Pulsar Timing Array and in the future with the
Square Kilometre Array SKA (see figure~\ref{fig:gw_spectrum}).

Other cosmological implications of a first-order QCD phase transition
are that the cold dark matter density is diluted by the same factor as
the baryon number density, i.e.\ by up to a factor $10^{-9}$. To
explain the present cold dark matter relic density within the standard
WIMP scenario, one needs a different WIMP annihilation cross section
$\sigma_{\rm ann}$ as the cold dark matter density scales as
$\Omega_{\rm CDM} \sim \sigma_{\rm weak}/\sigma_{\rm ann}$ or by a
substantially larger WIMP mass. Both cases can be probed by the LHC!
The large-scale structure power spectrum responsible for galaxy and
galaxy cluster formation would be modified up to a mass scale of $M
\sim 10^9 M_\odot$ which is the typical mass scale of dwarf galaxies.
Without QCD inflation, effects can be only up to the horizon mass $\sim
10^{-9}M_\odot$ at the phase transition point.  Also, seeds of
(extra)galactic magnetic fields can be generated by charged bubble
collisions during the phase transition. The observed magnetic field
today in our galaxy is $B\sim 10^{-5}$~G, extragalactic ones are in
the range of $B\sim 10^{-7}$~G. One needs primordial seed fields of
$B=10^{-30}\dots 10^{-10}$~G.  As the electroweak phase transition is
a crossover for large Higgs masses, a first order QCD phase transition
in the early universe would provide then a possible explanation of
those cosmological magnetic fields within the standard model again (see
\cite{Caprini:2009yp} for a detailed calculation).


\ack 

This work is supported by BMBF under grant FKZ 06HD9127, by DFG under
grant PA1780/2-1 and within the framework of the excellence
initiative through the Heidelberg Graduate School of Fundamental
Physics, the Gesellschaft f\"ur Schwerionenforschung GSI Darmstadt,
the Helmholtz Research School for Quark Matter Studies, the Helmholtz
Graduate School for Heavy-Ion Research (HGS-HIRe), the Graduate Program
for Hadron and Ion Research (GP-HIR), the Helmholtz Alliance Program
of the Helmholtz Association, contract HA-216 "Extremes of Density and
Temperature: Cosmic Matter in the Laboratory", and CompStar, a
research networking program of the European Science Foundation.


\section*{References}

\bibliographystyle{utphys}
\bibliography{all,literat}

\end{document}